\pdfoutput=1

 \documentclass[preprint2]{aastex}
 \usepackage{spr-astr-addons} 
 \usepackage{eurosym}
 \usepackage{hyperref}
 \usepackage[3D]{movie15}

\begin{document}

\title{Augmented Reality in Astrophysics}

\shorttitle{Augmented Reality in Astrophysics}
\shortauthors{Vogt and Shingles}

\author{Fr\'ed\'eric P.A. Vogt\altaffilmark{1,2} and Luke J. Shingles\altaffilmark{1}}
\email{fvogt@mso.anu.edu.au}
\altaffiltext{1}{Research School of Astronomy and Astrophysics, Australian National University, Cotter Road, Weston Creek, ACT 2611, Australia.}
\altaffiltext{2}{Contact: fvogt@mso.anu.edu.au}

\begin{abstract}
Augmented Reality consists of merging live images with virtual layers of information. The rapid growth in the popularity of smartphones and tablets over recent years has provided a large base of potential users of Augmented Reality technology, and virtual layers of information can now be attached to a wide variety of physical objects. In this article, we explore the potential of Augmented Reality for astrophysical research with two distinct experiments: (1) \emph{Augmented Posters} and (2) \emph{Augmented Articles}. We demonstrate that the emerging technology of Augmented Reality can already be used and implemented without expert knowledge using currently available apps. Our experiments highlight the potential of Augmented Reality to improve the communication of scientific results in the field of astrophysics. We also present feedback gathered from the Australian astrophysics community that reveals evidence of some interest in this technology by astronomers who experimented with Augmented Posters. In addition, we discuss possible future trends for Augmented Reality applications in astrophysics, and explore the current limitations associated with the technology. This Augmented Article, the first of its kind, is designed to allow the reader to directly experiment with this technology.
\end{abstract}

\keywords{Data Analysis and Techniques -- Tutorial}

\section{Introduction}\label{Sec:intro}
The market for camera-equipped smartphones and tablets has been rapidly expanding over the past few years. Although not marketed as a primary use-case, these widespread devices have all of the hardware required to enable Augmented Reality (AR) applications \citep[][]{Papagi08}. We define AR as the combination of live images with virtual layers of additional content. Our definition follows \cite{Azuma97}, with the exception that we do not restrict the additional content to 3D models, but also include other types of content, such as images, sounds, and videos. The principle of AR, in which virtual content is added on top of a real environment, is not to be confused with Virtual Reality, where the environment is mostly or totally virtual \citep[][]{Milgram94,Azuma01}. There exist two distinct types of AR, which differ in the way that the virtual layer associated with a given environment is identified: location-based AR and image-based AR \citep[][]{Cheng12}. In the case of location-based AR, applications rely on the spatial position and orientation of the device to select and display location-relevant information. For image-based AR, applications use image recognition algorithms to trigger the display of relevant content over a recognised physical pattern. In this article, we focus on image-based AR applications. NASA's \emph{Spacecraft 3D} is one example of such an application, where users can overlay 3D models of various spacecraft on top of a specific target image\footnote{ The Spacecraft 3D app is free to download. See NASA's website for more information. \href{http://www.nasa.gov/mission\_pages/msl/news/app20120711.html}{http://www.nasa.gov/mission\_pages/msl/news/app20120711.html}}. AR is not restricted to smartphones and tablets, but is also compatible with a wider range of hardware, such as head-mounted and head-up displays. At the moment, smartphones and tablets vastly outnumber other AR-capable devices. \cite{Azuma97}, \cite{Azuma01}, and \cite{Krevelen10} present extensive reviews of the history of AR and associated technologies. 

The potential of AR for educational and teaching purposes in science and engineering has already been identified \citep[see, e.g.,][]{Martin11}. Recent examples include a real-time visualisation system of magnetic fields \citep[][]{Matsumoto12}, and an augmented electrical engineering laboratory for distance learning \citep[][]{Mejias12}. Several educational AR applications have also been proposed in the field of astrophysics. \cite{Sin09} developed an AR system relying on head-mounted displays, allowing students to interact with a 3D model of the Solar system. \cite{Shelton02} describe another example of an AR device illustrating the Sun-Earth interaction. And in the field of archeo-astronomy, \cite{Schiavo09} advocates for the implementation of an AR system allowing the inclusion of ancient buildings within our modern world to better transmit research results to non-specialists. A more extended list of educational AR applications in various fields of science is presented in \cite{Cheng12}, to which we refer the reader for further details.

In this article, we focus on implementations of AR for astrophysical research. We put forth the view that AR may have a great but yet under-exploited potential in astrophysics. We have identified two specific cases in which AR may improve how scientific results are communicated between researchers. These are (1) \emph{Augmented Posters} in scientific conferences, and (2) \emph{Augmented Articles} in scientific journals. The underlying motivation behind the use of AR at the research level is to develop new ways to share multi-dimensional datasets. Whether numerical 3D simulations or Integral Field Spectroscopy (IFS) observations, modern astronomers are confronted with multi-dimensional data sets on a daily basis. The comprehension of complex data sets often benefits from alternative visualisation methods, such as interactive 3D models or animations. Enabling the wide dissemination of such animations between scientists can therefore improve the communication of scientific results within the community.

Scientific research is shared in three main ways: articles, talks and posters. Until recently, neither articles nor posters were suitable for publishing interactive content. However, documents in \emph{Adobe Portable Document Format} (PDF) are now able to contain animated 3D models. This technology has a strong potential for astrophysics \citep[][]{Barnes08}, especially since all astrophysical journals are now published online. AR is another technology that has the potential to add more flexibility to scientific articles and posters. Yet, because it has not yet reached the \emph{Plateau of Productivity} in the so-called hype cycle \citep[see, e.g.,][]{Fenn08}, the development of AR technology, and more specifically its potential wide-spread usage in science, may be strongly influenced by pre-conceived (positive or negative) opinions from researchers, who may not have had the opportunity to directly experiment with AR. 

In this article, we address these issues directly. First, we have implemented Augmented Poster sessions at the Astronomical Society of Australia (ASA) Annual General Meeting (AGM) 2012 in Sydney. This conference, gathering astronomers from all academic levels and from all fields of theory, observations and instrumentation, is a perfect opportunity to test AR technology on a representative sample of astronomers in Australia. In the first part of this article, we present the outcome of this experiment. We analyze the usage of AR during the conference and the feedback gathered from the conference attendees. 

Secondly, this article has been \emph{augmented} in two different ways, so that the readers can directly experiment with AR for themselves, discover some of its potential and forge their own opinion regarding the possible usage of AR at the research level. Our experiments rely on two distinct apps and their associated internet databases: \emph{Shortcut} and \emph{Layar}. Both these apps are free to download for iOS and Android devices (as well as for Windows phones in the case of Shortcut). This article is set up to be compatible with both systems, and the reader is encouraged to experiment with the apps. The concept of AR is very general and is not restricted to these two proprietary apps. At this time however, the Shortcut and Layar apps are perfect to illustrate the concept and potential of AR for astrophysical research.

This article is organised as follows. We briefly discuss the notion of image recognition in Section~\ref{sec:recognition}. We describe our experiment with Augmented Posters in Section~\ref{Sec:posters}. We introduce the concept of Augmented Articles and illustrate it with a practical example in Section~\ref{sec:papers}. We discuss potential directions for the expansion of AR in the field of astrophysics in Section~\ref{sec:future}, and explore current limitations of AR for astrophysical research in Section~\ref{sec:limitations}. Finally, our conclusions are in Section~\ref{Sec:summary}. The ethical aspects of this research have been approved by the Human Research Ethic Committee of the Australian National University (protocol 2013/201).

\section{Image recognition and QR codes}\label{sec:recognition}

Image-based AR requires that physical objects can be successfully identified from imaging data, in order to select and display a relevant virtual layer of information. Research into computer vision is ongoing, and significant advances have been made in image recognition algorithms. In addition, the computing power available to smartphones and tablets has jumped significantly with the transition from local-computing to a connected cloud-computing model in which extremely intensive processing tasks are offloaded to powerful computers in remote data centers.

\begin{figure}[hbt!]
\centerline{\includegraphics[scale=0.25]{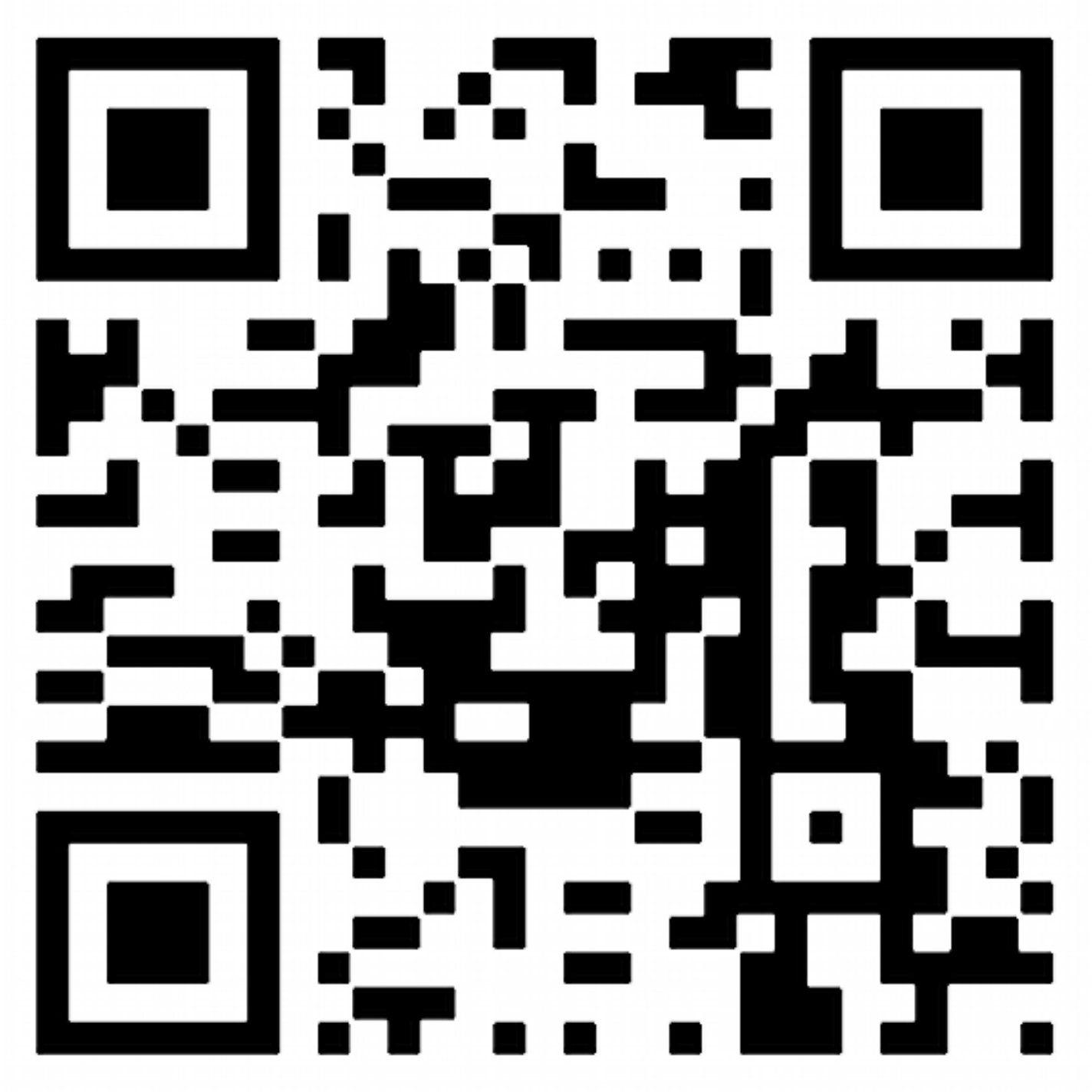}\qquad\includegraphics[scale=0.25]{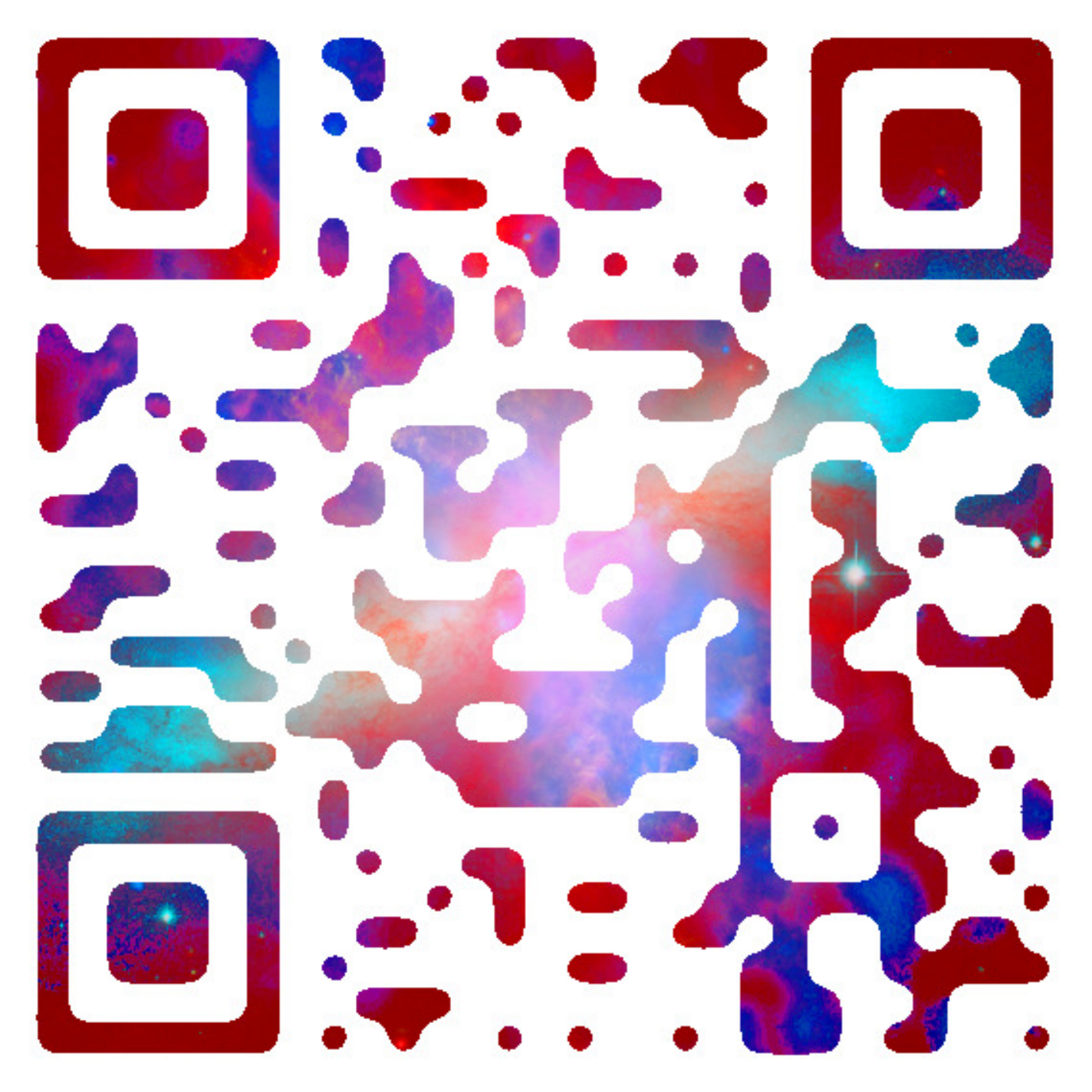}}
\caption{Two QR codes pointing to the same URL (\href{http://hdl.handle.net/102.100.100/9788}{http://hdl.handle.net/102.100.100/9788}), but with distinct designs. M82 image credits : X-ray: NASA/CXC/JHU/D.Strickland; Optical: NASA/ESA/STScI/AURA/The Hubble Heritage Team; IR: NASA/JPL-Caltech/Univ. of AZ/C. Engelbracht}\label{fig:QR}
\end{figure}

Before these advances, image recognition has historically not been a practical solution for the machine identification of physical objects. Automated object recognition has typically been (and often still is) aided by the attachment of machine-readable patterns (e.g. barcodes) to physical items. The image of a barcode associated with an object can be easily decoded, and requires only minimal computing power. Quick-response codes  \citep[QR codes, see, e.g.,][p.~341]{Fuhrt11} are two-dimensional barcodes, originally developed in the automobile industry. Over recent years, the range of applications of QR codes has significantly widened to reach the general public, for example in magazines, adverts and on product labels. Today, many smartphones and tablets are capable of reading QR codes. Two different examples of QR codes are shown in Figure~\ref{fig:QR}. In addition to the original black and white design, QR code readers are usually compatible with some visual modifications, such as smooth edges or coloured content, provided that the contrast is kept high.

The advances in image recognition technology have coincided with the emergence of AR-capable devices. As a result, most implementations of image-based AR recognise objects using computer vision algorithms. In principle, QR codes can be used to achieve the same outcome, i.e. the identification of the appropriate virtual information layer. But as stand-alone features, QR codes are not a complete AR system as per our definition in Section~\ref{Sec:intro}.

QR codes have come to be well known and easily recognisable marks. This is a strong advantage, because many users immediately know that additional content is available with a QR code reader. By comparison, AR apps that rely on general-purpose image recognition must use other means to signal the availability of additional content, since users typically only open an AR app after identifying a compatible document.

However, general-purpose image recognition has some benefits over the use of QR codes. For example, while QR codes must be incorporated at the design stage of print media, general-purpose image recognition enables existing documents to be associated with new layers of information, even after publication. Prompting users to open an AR app is typically achieved via text instructions or the presence of an app logo in the document, but could alternatively, with general-purpose image recognition, be solely external to the document itself, e.g., at a conference that advertises AR-enabled poster sessions (see Section~\ref{Sec:posters}).

A discussion of which method (QR codes or general-purpose image recognition) is best for AR applications is outside the scope of this paper. In fact, the technologies can be used together as, for example, in the case of the Layar app (see Section~\ref{sec:papers}). The two experiments we discuss in Sections~\ref{Sec:posters} and \ref{sec:papers} solely rely on general-purpose image recognition, and not on QR codes.

\section{Augmented Posters}\label{Sec:posters}

\subsection{Description of the experiment}
Shortcut\footnote{ \href{http://shortcutmedia.com/}{http://shortcutmedia.com/}} is an app that allows users to scan a compatible document with their smartphone or tablet to access additional linked material. The app is mostly used with newspapers worldwide to access online references or videos, and with advertising companies on billboards. Shortcut is not an AR app as per our definition (see Section~\ref{Sec:intro}), since it doesn't merge virtual data sets with a live camera stream, but rather with a still image. In that sense, the app is used in a similar way to QR codes, but with distinct differences. With Shortcut, publishers are not limited to a single link per scan, but can point to several URLs or documents from the same target, without having to manually set up any external webpage or online storage space. Because it relies on general-purpose image recognition technology, the app does not require the addition of any specific mark to the underlying document (although the compatibility of a document with Shortcut ought to be indicated graphically for rapid identification by the users, see Section~\ref{sec:recognition}). Finally, the app provides a uniform structure to access and visualize virtual layers of information for all supported documents. These three features place the Shortcut app a step closer to being a \emph{real} AR app as defined in Section~\ref{Sec:intro} (see Section~\ref{sec:papers} for an example) than most QR-based solutions.

In a scientific conference, this app works as follows. Poster authors upload their poster to an online database, and fill the associated virtual layer of information via a simple upload page. The content of the virtual layer can be very diverse, and may contain (for example) an email address, a URL and/or a PDF document. By scanning a poster of interest, conference attendees can access within a few seconds the virtual layer on their smartphone, send it via email, or share it on social media directly from within the app. Image recognition is fast, but requires internet access. It can be expected that most conferences will have wireless internet available during the poster sessions. Shortcut however does not require internet access on-the-spot. Poster snapshots can be saved and the database searched at a later time.

Augmented Poster sessions were implemented at the ASA AGM 2012. The ASA AGM is the yearly gathering of astronomers based in Australia. The conference covers all Australian astronomical research areas, spanning multiple wavelength bands from the radio to the X-ray, in all theoretical, observational and instrumentation domains. The ASA AGM 2012 was hosted by the University of New South Wales in Sydney from the 1st-6th July 2012. It comprised 87 talks spread over the 4.5 days of the conference, with 254 registered attendees and 108 posters. The ASA AGM is well suited for this first large scale implementation of the AR-like Shortcut technology in a conference because it gathers astronomers from a wide range of backgrounds and academic levels. 

\begin{figure}[htb]
\centerline{\includegraphics[scale=0.25]{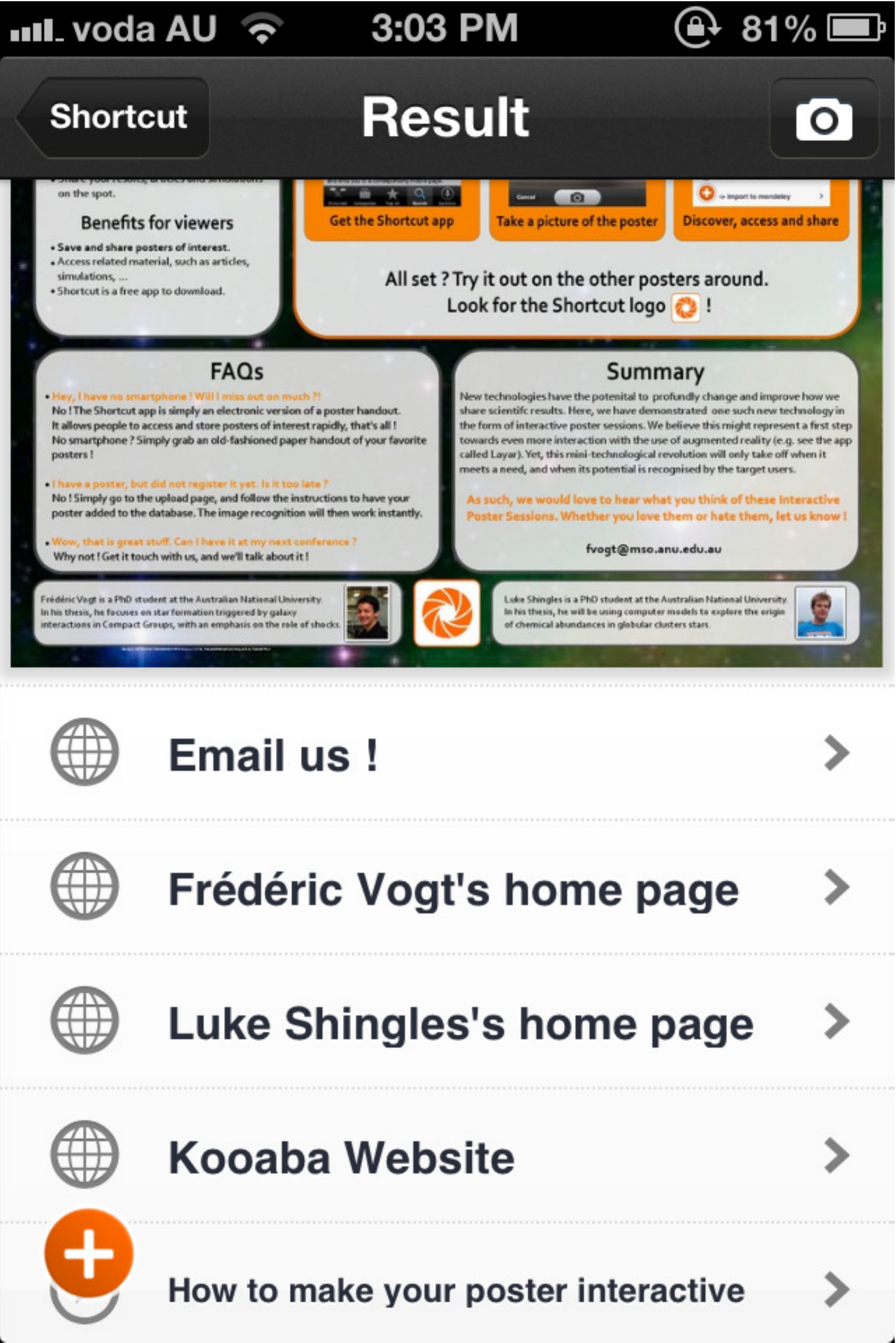}}
\caption{Screen capture of the virtual layer of information attached to the poster in Appendix~\ref{app:poster}, accessible by snapping a picture using a smartphone or tablet with the Shortcut app.}\label{fig:shortcut}
\end{figure}

We submitted a poster at the ASA AGM explaining the principle of Augmented Poster sessions and how to access the virtual layers of information. A copy of this poster is available in Appendix~\ref{app:poster}. At the time of publication, the poster is still compatible with the Shortcut app. The poster is expected to remain active and compatible for the foreseeable future (see Section~\ref{sec:limitations} for a discussion on why this may not be the case forever). It is therefore possible to access the virtual layer of information available to conference attendees by taking a picture of the image in the appendix with a smartphone and the Shortcut app. A screen capture of the virtual layer of information is shown in Figure~\ref{fig:shortcut}.

\subsection{Experiment outcome}\label{Sec:results}
\subsubsection{The usage of Shortcut during the conference}\label{Sec:usage}

\begin{figure*}[htb!]
\centerline{\includegraphics[scale=0.4]{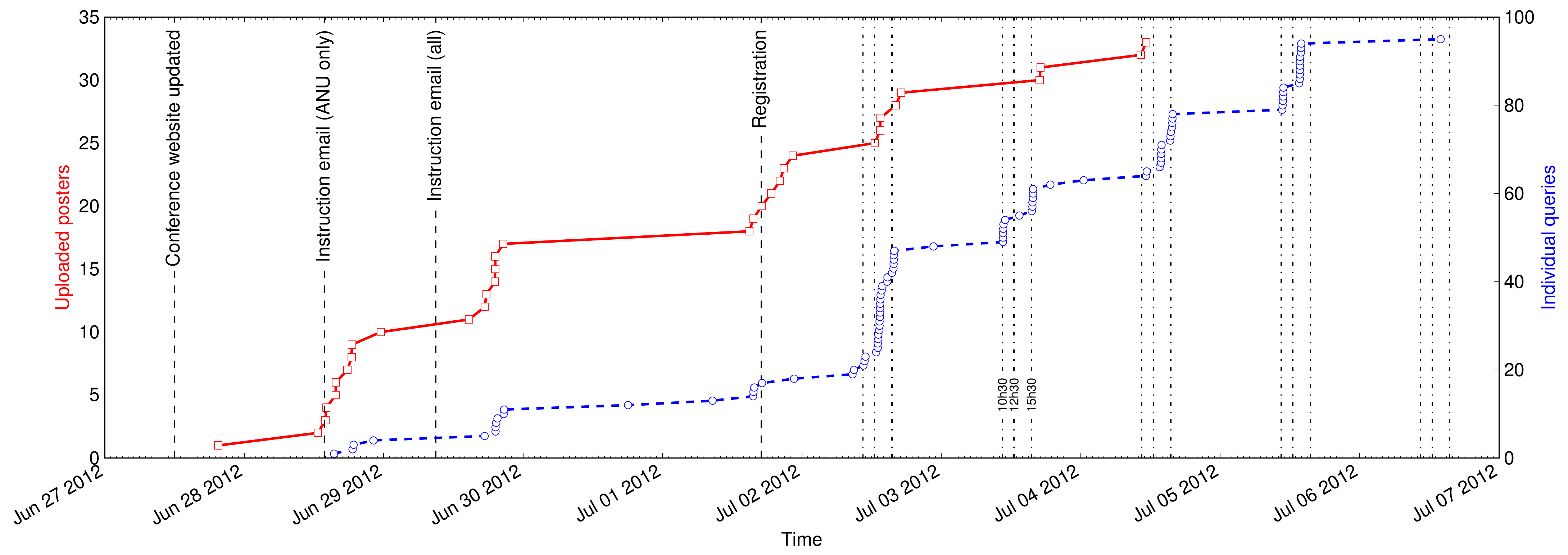}}
\caption{Time evolution of the number of poster uploads to the database (solid red curve, square markers), and individual successful queries to the database via the Shortcut app (dashed blue curve, circle markers). Each data point corresponds to one upload or query. Vertical dashed lines indicate particular events occurring prior to the conference. Dot-dashed lines denote the time of the morning, noon and afternoon breaks on each day of the conference. }\label{Fig:uploads}
\end{figure*}

We tracked each augmented poster during the conference using the Shortcut administration website. The system is anonymous and does not allow the identification of the app users. A total of 98 database searches were recorded, spread among 33 posters uploaded to the database. In Figure~\ref{Fig:uploads}, we show the timing of each of the 33 poster uploads to the database (solid red line). We also show the timing of every individual search (dashed blue line). Only successful searches are recorded. 

Poster uploads are clustered around specific events. A first peak occurred after we sent an email to astronomers at the Australian National University (ANU) and encouraged them to upload their poster to the database. A second group of uploads occurred as the general upload instruction email was sent to all poster authors attending the ASA AGM. Finally, two additional upload events occurred at the conference registration and after the first poster session. 

About 50\% of all the Augmented Posters were uploaded to the database on or after the conference registration. This is most likely the consequence of a lack of clear communication with poster authors prior to the beginning of the conference. Many conference attendees first discovered the Augmented Poster technology at the conference and only then decided to upload their poster to the database. With a total of 33 Augmented Posters ($\sim$30\% of all posters), the concept was nevertheless well supported by poster authors overall given the novelty of the technology.

\begin{figure}[htb!]
\centerline{\includegraphics[scale=0.45]{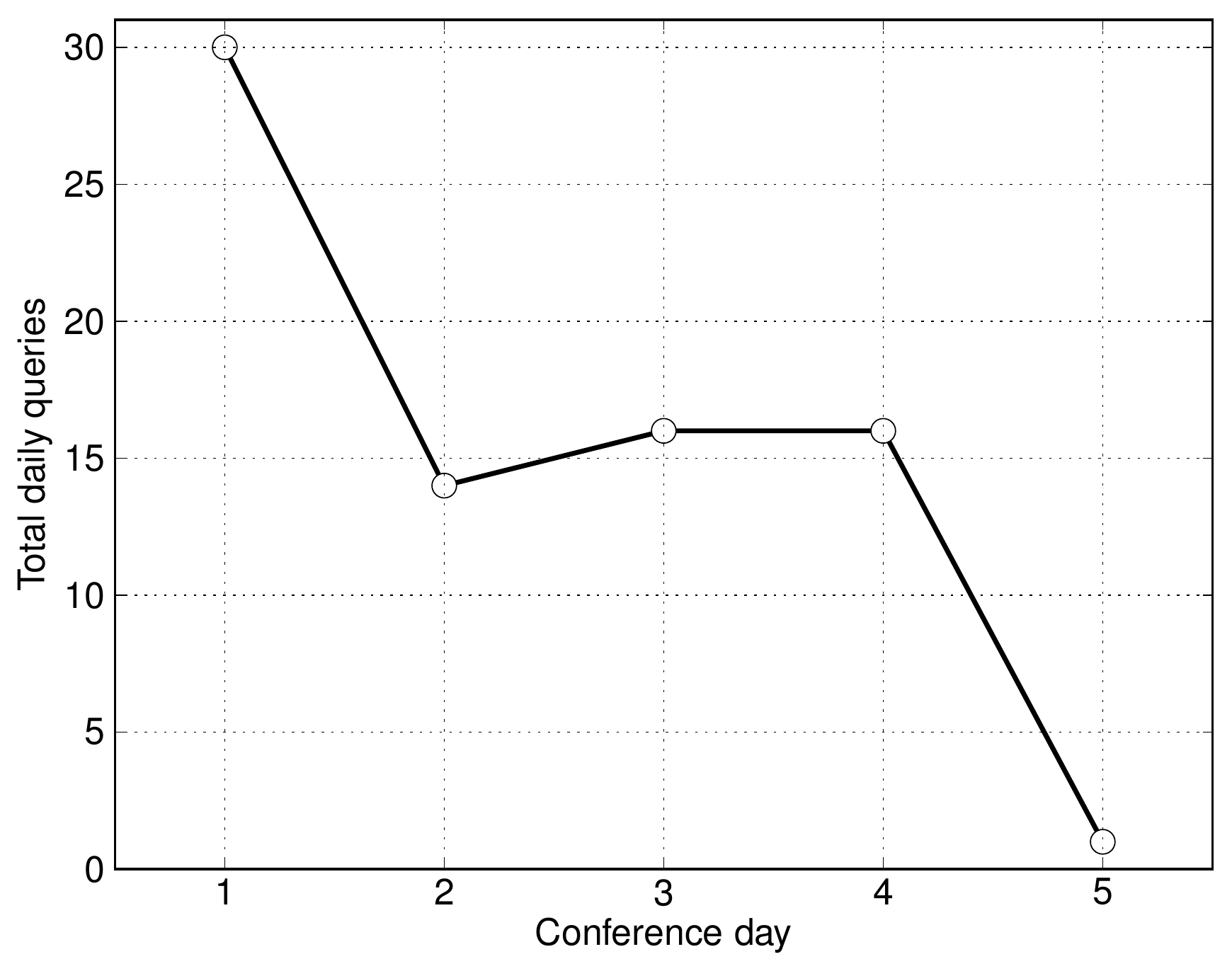}}
\caption{Total daily queries as a function of the conference day. After an initial burst of interest, the usage of the app stabilizes during the week at $\sim$15 queries per day. }\label{Fig:usage}
\end{figure}

The first poster searches occurring prior to the conference opening are due to poster authors testing the Shortcut app on their own poster. The queries during the conference occurred almost exclusively during the three daily breaks in the conference schedule. For clarity, we show in Figure~\ref{Fig:usage} the total number of queries on each conference day. After an initial peak on the first day with 30 hits, the other days have all had a stable number of $\sim$15 hits each, the final conference day excepted. 

\begin{figure}[htb!]
\centerline{\includegraphics[scale=0.48]{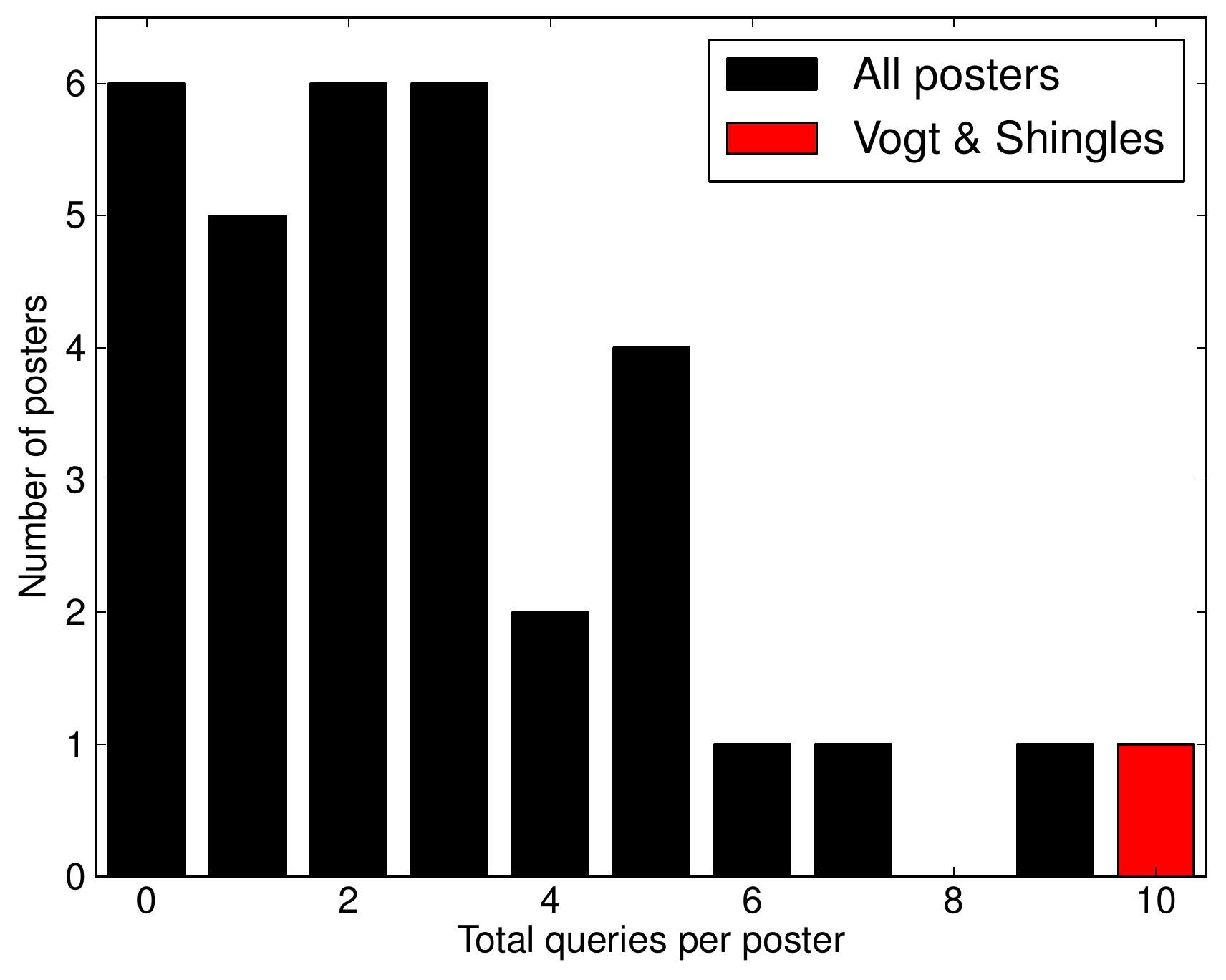}}
\caption{Distribution of the total number of successful queries per poster. The red bar corresponds to our poster explaining the concept of Augmented Poster sessions. }\label{Fig:popularity}
\end{figure}

In Figure~\ref{Fig:popularity}, we show the number of Augmented Posters as a function of the number of successful searches received before and during the conference. Our poster (see Appendix~\ref{app:poster}) dedicated to the concept of Augmented Posters, is at the top of the list with a total of 10 hits over the entire week. We have no direct measurement of how many individual conference attendees used the Shortcut app. We estimate that around $\sim$20 people did, of which 5-10 used the app multiple times. These estimates rely on the average number of daily hits, and on in-situ observations during the conference itself. With  $<$10\% of all attendees trying the app and 2-5\% using it at multiple occasions, the uptake of AR technology in the form of Augmented Posters was 3-15 times less for conference attendees in general compared to poster authors. This trend suggests that poster authors are more keen to experiment with new technologies than conference attendees in general, in the hope that it may give more visibility to their poster and their science during the conference.

\begin{figure}[htb!]
\centerline{\includegraphics[scale=0.47]{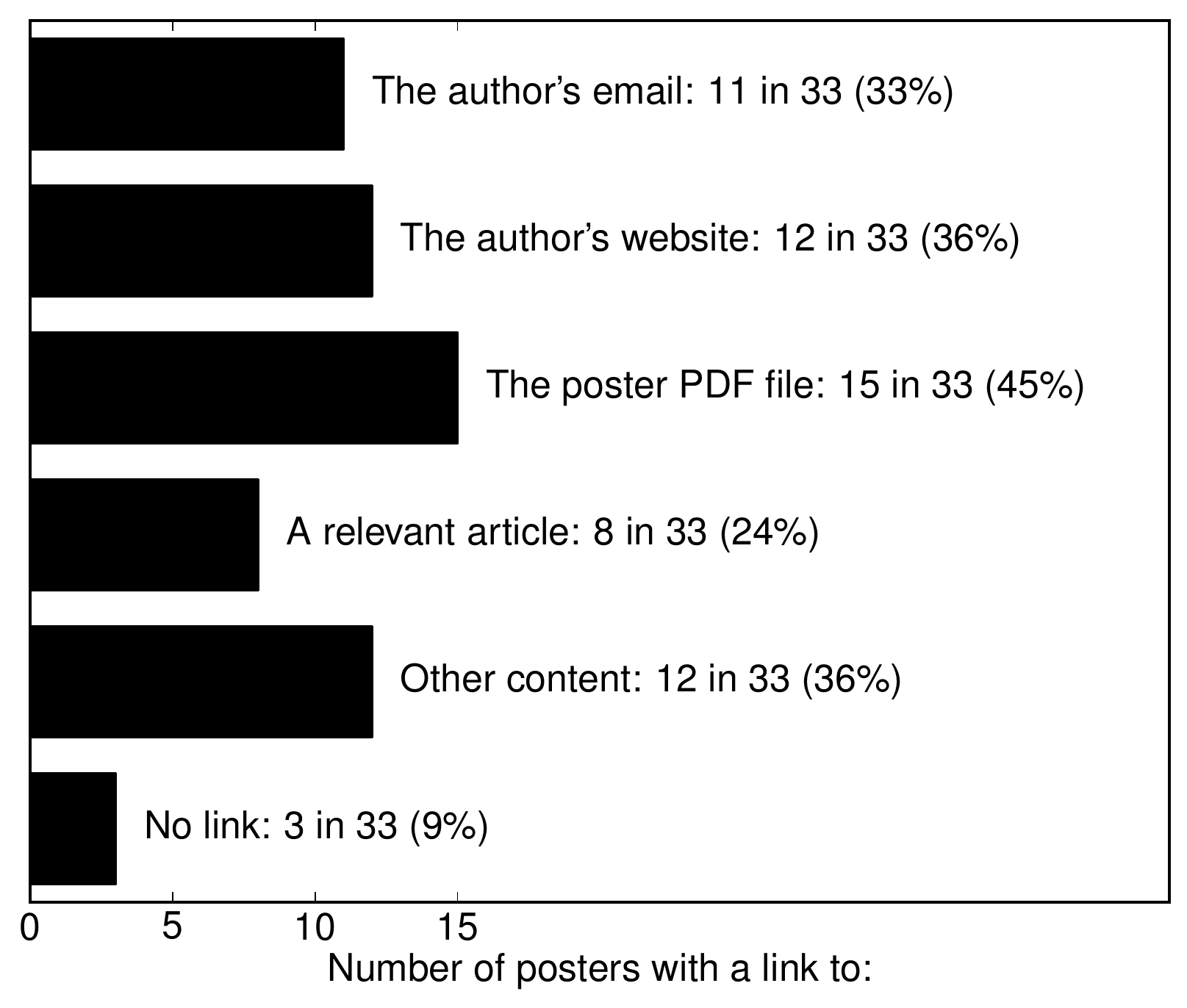}}
\caption{Number of posters uploaded to the database with links to the author's email, the author's website, a copy of the poster in PDF, a related article, or other type of content (e.g. other types of websites, videos, etc). Three posters have been uploaded to the database with an empty virtual layer of information. }\label{Fig:links}
\end{figure}

No strict rules were given to poster authors regarding the content of the virtual layer of information. The instruction email contained suggestions, but did not attempt to impose any restrictions. As a result, the different virtual layers of information of the 33 Augmented Posters contained a wide range of information. In Figure~\ref{Fig:links}, we show the number of Augmented Posters linking to different types of content. 

A PDF copy of the poster was, with 15 links, the most popular item by a short lead. We note that three posters were uploaded to the database with an empty virtual layer of information. This is most certainly another signature of communication issues with poster authors prior to the conference. The small uptake of Augmented Posters by conference attendees might be a consequence of the eclectic type of material accessible. It might be beneficial in future implementations of similar Augmented Poster sessions to request poster authors to populate their virtual layer of information with some minimal information, such as their contact details and a copy of the poster, for example. However, leaving some degrees of freedom might also have a positive impact, as some conference attendees pointed out that discovering what every virtual layer of information contained motivated them to use the app on multiple occasions.

\subsubsection{Feedback from the conference attendees}\label{Sec:feedback}

A link to an online feedback survey was sent to all attendees of the ASA AGM a few days after the end of the conference. The survey contained 9 questions, and typically took less than a few minutes to be filled. The complete survey (questions and multiple-choice answers) is located in Appendix~\ref{App:survey}. This survey was designed as an internal audit tool. No experienced social statistician was involved in the construction (or analysis) of the questions. We received a total of 18 replies ($\sim$7\% of all conference attendees), 17 from astronomers who attended the conference, and 1 from an astronomer who submitted a poster but did not attend the conference. The small number statistics we are dealing with coupled with the original \emph{internal audit} nature of the survey prevent us from performing an extensive analysis of all the data gathered. Yet, we believe some of the data is worth discussing here. 

The responses came from astronomers of all academic levels: 2 undergraduate students, 10 graduate students, 4 researchers in a post-doctoral position and 2 professors. Among these 18 people,  8 used the Shortcut app during the conference, 11 submitted a poster at the ASA AGM, and 8 out of these 11 posters were uploaded to the Shortcut database. Among the respondents who did not use the Shortcut app, 2 said they were not interested, 5 said they had no smartphone, and 3 had other reasons.

\begin{figure}[htb!]
\centerline{\includegraphics[scale=0.47]{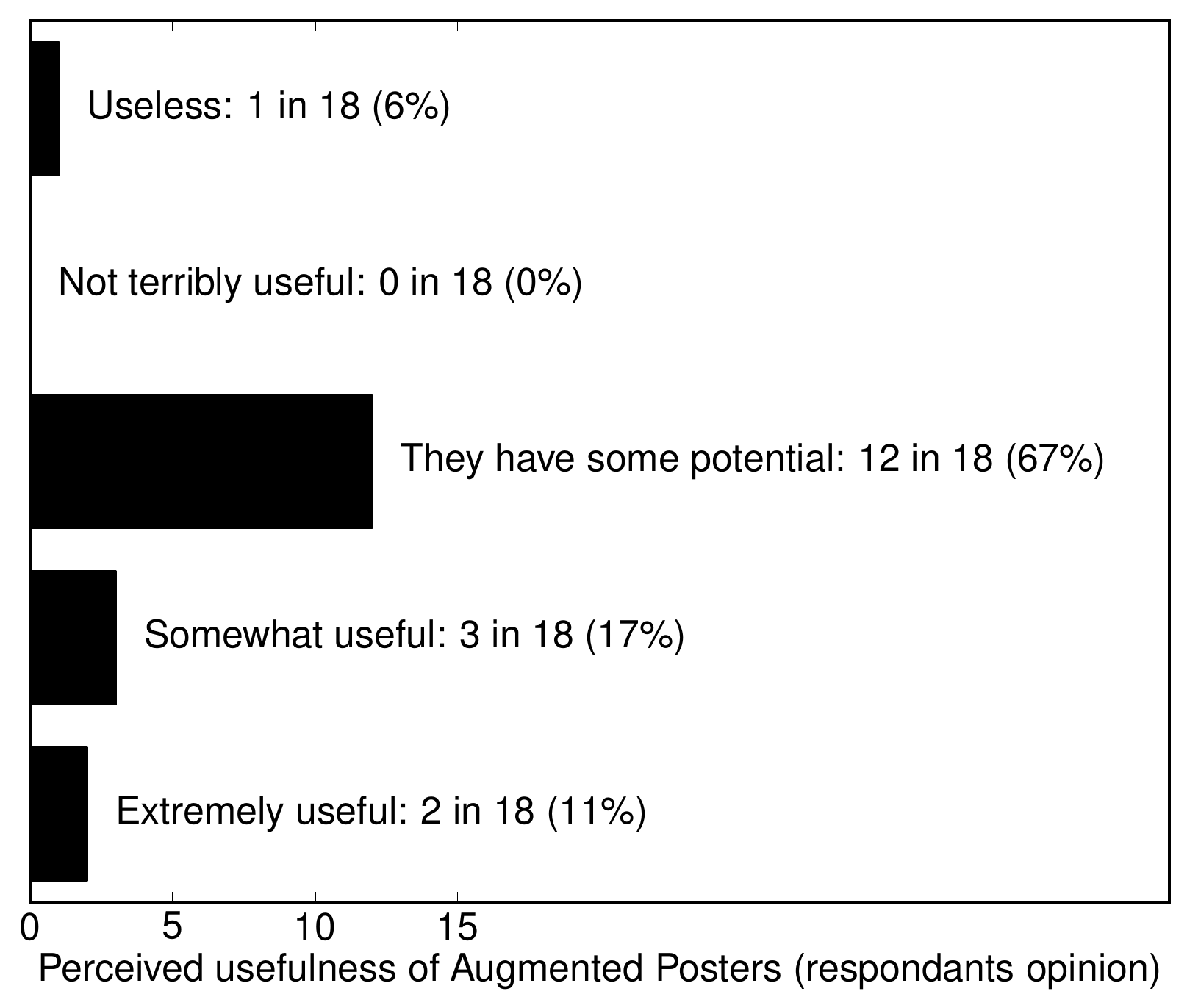}}
\caption{Distribution of answers regarding the perceived usefulness of Augmented Poster sessions. }\label{fig:potential}
\end{figure}

In the feedback survey, we also asked respondents to rank the perceived usefulness of Augmented Poster sessions on a scale from 1 (\emph{Useless}) to 5 (\emph{Extremely useful}). The distribution of answers is shown in Figure~\ref{fig:potential}. Except for one very negative opinion, all responses are either neutral or positive. In addition, 15 persons (83\%) who answered our survey would be interested to see Augmented Poster sessions implemented again at future conferences. 

It is clear that we are here strongly biased towards people that used the Shortcut app during the conference. Hence, our survey suggests that astronomers who used the Shortcut app are globally supportive of the idea of more implementations of Augmented Poster sessions in astrophysical conferences. These future implementations would give more astronomers the opportunity to discover and get acquainted with the concept and technology of AR.

\section{Augmented Articles}\label{sec:papers}

We now present a different research-level application of AR in astrophysics, in the form of Augmented Articles. The main feature driving this experiment is the ability to easily share animations and videos in printed materials. By taking a picture of a given page in a given augmented article, readers can directly view videos and other related material on their smartphone. One aspect of this technology is that it allows access to additional online material without the need for a computer. This can be important for developing countries, where smartphones outnumber computers, or in situations when carrying a computer would be impractical. A similar approach was adopted by the journal \emph{Neurosurgery} in 2011 by implementing widespread usage of QR codes in their articles \citep[][]{MacRae11}. In the next section, we implement an example of an Augmented Article for the field of astrophysics (the first of its kind) to demonstrate the technology and its potential. In contrast with the AR-like Augmented Posters described previously, the concept of Augmented Articles illustrated in this section matches our definition of AR (see Section~\ref{Sec:intro}) by merging live images with virtual data sets. 

\subsection{The animated 3D structure of SNR N132D}\label{sec:n132d}

SNR N132D is a young supernova remnant located in the Large Magellanic Cloud (LMC). \cite{Vogt11} used the Wide Field Spectrograph \citep[WiFeS,][]{Dopita07, Dopita10} at Siding Spring Observatory to map the oxygen-rich ejecta in this system. The velocity of the oxygen-rich knots, expelled during the supernova explosion, are identified by studying the red- and blueshift of the [O~{\footnotesize III}] $\lambda$5007 {\AA} forbidden emission line. The initial data cube axes (X [arcsec], Y [arcsec], $\lambda$ [\AA]) are then transformed to (X [arcsec], Y [arcsec], v$_\mathrm{r}$ [km s$^{-1}$]), where $v_\mathrm{r}$ is the radial velocity of the ejecta. Assuming free expansion of the ejecta, a distance to the LMC of 50 kpc \citep[][]{vandenBergh99}, and an age of $\sim$2500 years, the radial velocity axis is converted to a spatial dimension. Thus, they obtain an accurate 3D spatial map with axes (X [pc], Y [pc], Z [pc]) of the oxygen-rich filaments in SNR N132D.

\begin{figure*}[htb!]
\centerline{\includegraphics[scale=0.2]{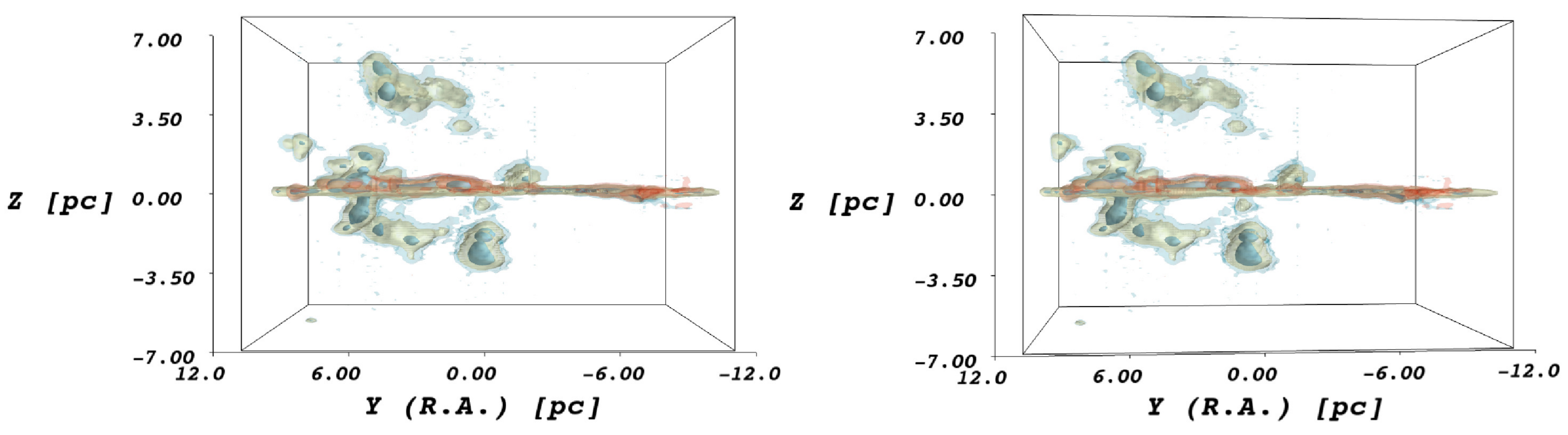}}
\caption{Cross-eyed stereo pair of the oxygen-rich ejecta in SNR N132D. The axes are in parsec. The ejecta form a distorted ring structure. See \cite{Vogt12} for detailed instruction on how to visualize stereo pairs. }\label{fig:stereo}
\end{figure*}

\begin{figure*}[htb!]
\centerline{\includemovie[poster, toolbar, 3Dc2c = 0 0 1, 3Droo= 57, 3Droll = 90, 3Dlights= CAD, 3Dbg = 35 35 35, label=dice, text={\includegraphics[scale=0.3]{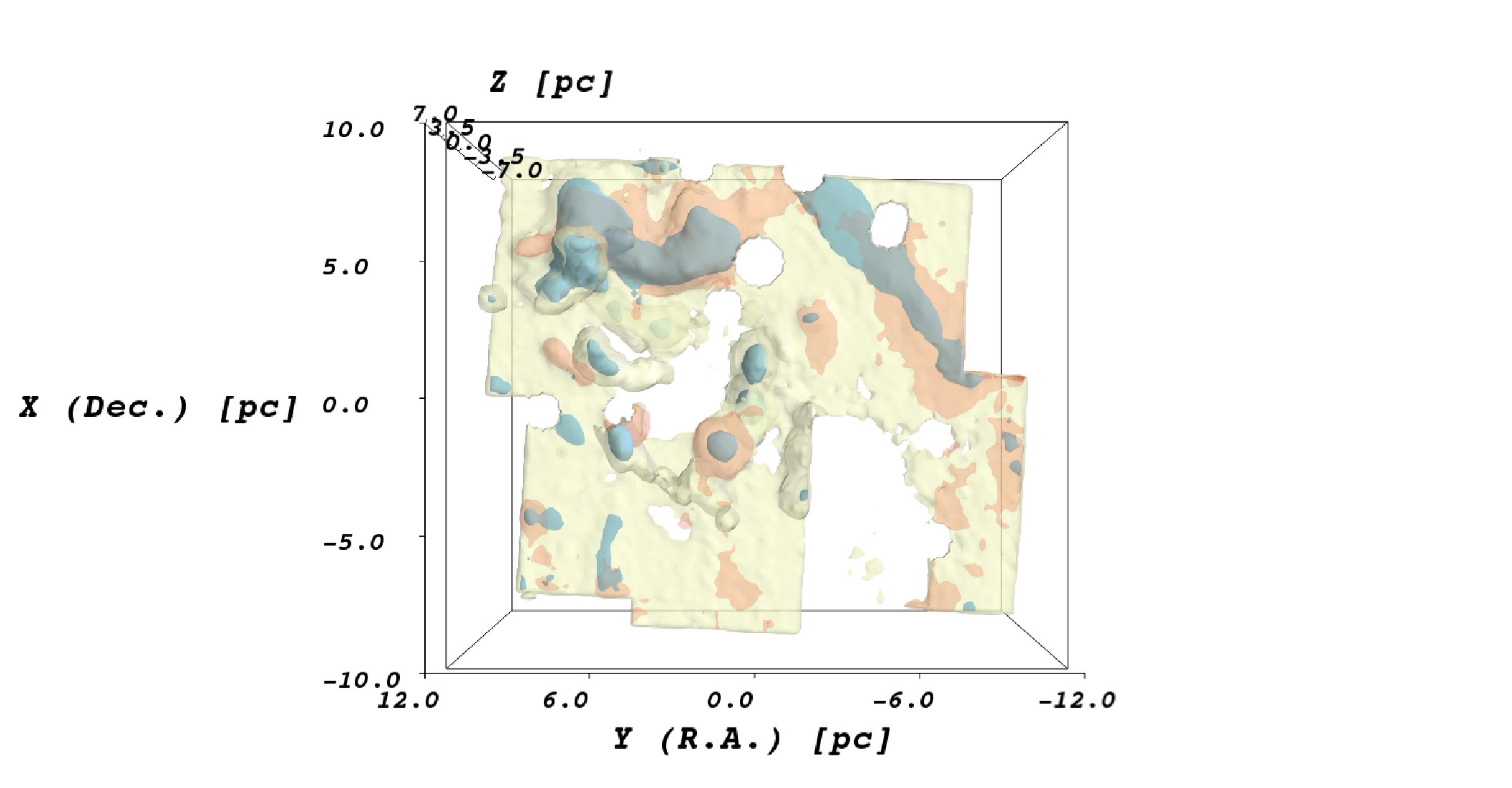}}]{0.98\linewidth}{ 0.45\linewidth}{./Vogt_fig9.u3d}}
\caption{3D projection of the oxygen-rich ejecta in SNR N132D, as seen from the Earth. The axes are in parsec. The ejecta form a distorted ring structure, best revealed in an animation of this 3D map. In the electronic version of this article uploaded on the arXiv server, an interactive 3D model can be loaded by clicking on the image (using Adobe Acrobat Reader to open the article). In addition, we also attach a virtual layer of information to this image, accessible with the Layar app, and containing a small animation of the 3D map. Both the stand-alone interactive 3D map and the animation are directly and freely accessible online (\href{http://dx.doi.org/10.4225/13/5195C67DAE4DD}{doi:10.4225/13/5195C67DAE4DD}).}
\label{fig:n132d}
\end{figure*}

Projections of this 3D map are shown in Figure~\ref{fig:stereo} and \ref{fig:n132d}. \cite{Vogt11} relied on interactive 3D models, stereo pairs and different 2D projections to show that the ejecta form a distorted ring. Here, we have recreated their 3D map using the latest versions of the Python module \emph{Mayavi}, resulting in smoother, more aesthetic maps. The scientific content of the 3D map is identical to the maps presented in \cite{Vogt11}, with the exceptions that back- and foreground stars have been removed from the 3D map, and the zero-velocity plane is not cropped.

The 3D map is $~$24 pc $\times$ 20 pc $\times$ 15 pc in size, and contains three iso-surfaces (light-blue with 75\% transparency, yellow with 50\% transparency and dark-blue with 0\% transparency) corresponding to different intensities of [O~{\footnotesize III}] emission, and one iso-surface (red) corresponding to H$\beta$ emission. Oxygen-rich ejecta are identified as hydrogen-free knots of [O~{\footnotesize III}] emission. Regions containing both H$\beta$ and [O~{\footnotesize III}] emission correspond to shocked interstellar medium surrounding SNR N132D. We refer the reader to \cite{Vogt11} for the detailed analysis of these 3D maps. In Figure~\ref{fig:stereo}, we show a cross-eyed stereo pairs of the oxygen-rich ejecta in SNR N132D seen from within the plane of the sky. While the ring structure of the ejecta is easily identifiable, using a cross-eyed visualisation technique with this stereo pair will allow the reader to obtain a depth feeling for the structure. \cite{Vogt12} provide detailed explanations on how to efficiently visualize stereo pairs.

In Figure~\ref{fig:n132d}, we show a projection of the 3D map of SNR N132D as seen from the Earth. The content of the map is similar to the maps shown in Figure~\ref{fig:stereo}, with the exception that the faintest [O~{\footnotesize III}]  iso-surface (light-blue and 75\% transparency) has been removed for clarity. In the electronic version of this article uploaded to the arXiv preprint server, this Figure contains an interactive PDF layer with the 3D map of SNR N132D that can be accessed using the Adobe Acrobat Reader software. An augmented layer containing an animation showing a rotation around the 3D map is also attached to Figure~\ref{fig:n132d}. In the next two sections, we describe these two layers (interactive and augmented) individually. Both the interactive 3D map and the animation are stored online and freely accessible (\href{http://dx.doi.org/10.4225/13/5195C67DAE4DD}{doi:10.4225/13/5195C67DAE4DD}).

\subsubsection{The interactive layer of Figure~\ref{fig:n132d}}

Similarly to \cite{Vogt11} and following the suggestion of \cite{Barnes08}, we include in the electronic version of this article uploaded to the arXiv preprint server an interactive model of the 3D map of SNR N132D. The interactive model can be accessed by clicking on Figure~\ref{fig:n132d}. At this time, interactive PDF can only be visualized using \emph{Adobe Acrobat Reader v8.0} or above. Other PDF viewers, such as Apple's \emph{Preview}, will only display the still projection of the map. We relied on the commercial \emph{PDF3DReportGen} software to transform VRML cubes created with Python to u3d cubes compatible with {\LaTeX} and PDF documents. 

The interactive 3D map allows the reader to rotate, zoom and fly around the 3D model. Such a 3D map is a very efficient tool to reveal the 3D structure of the oxygen-rich ejecta SNR N132D. The map also allows the display (or not) of the several layers of emission described previously, giving the reader total control over what he/she wants to see. For example, black spheres in the interactive map showing the positions where field stars have been removed can be hidden to improve the clarity of the oxygen-rich filaments, or displayed to identify the cropped regions (vertical cylinders) in the data cube.

This interactive layer is designed to be accessed with a computer - and is included in this article to illustrate how recent technological developments can improve our capacity to share complex data sets. At this stage, accessing the interactive 3D model from a smartphone or a tablet can be rather inefficient, due to a lack of proper software support for interactive PDF. In the next Section, we discuss how AR can complement an interactive layer by providing alternative ways to access and visualize complex data sets from smartphones and tablets.

\subsubsection{The augmented layer of Figure~\ref{fig:n132d}}

\begin{figure}[htb!]
\centerline{\includegraphics[scale=0.19]{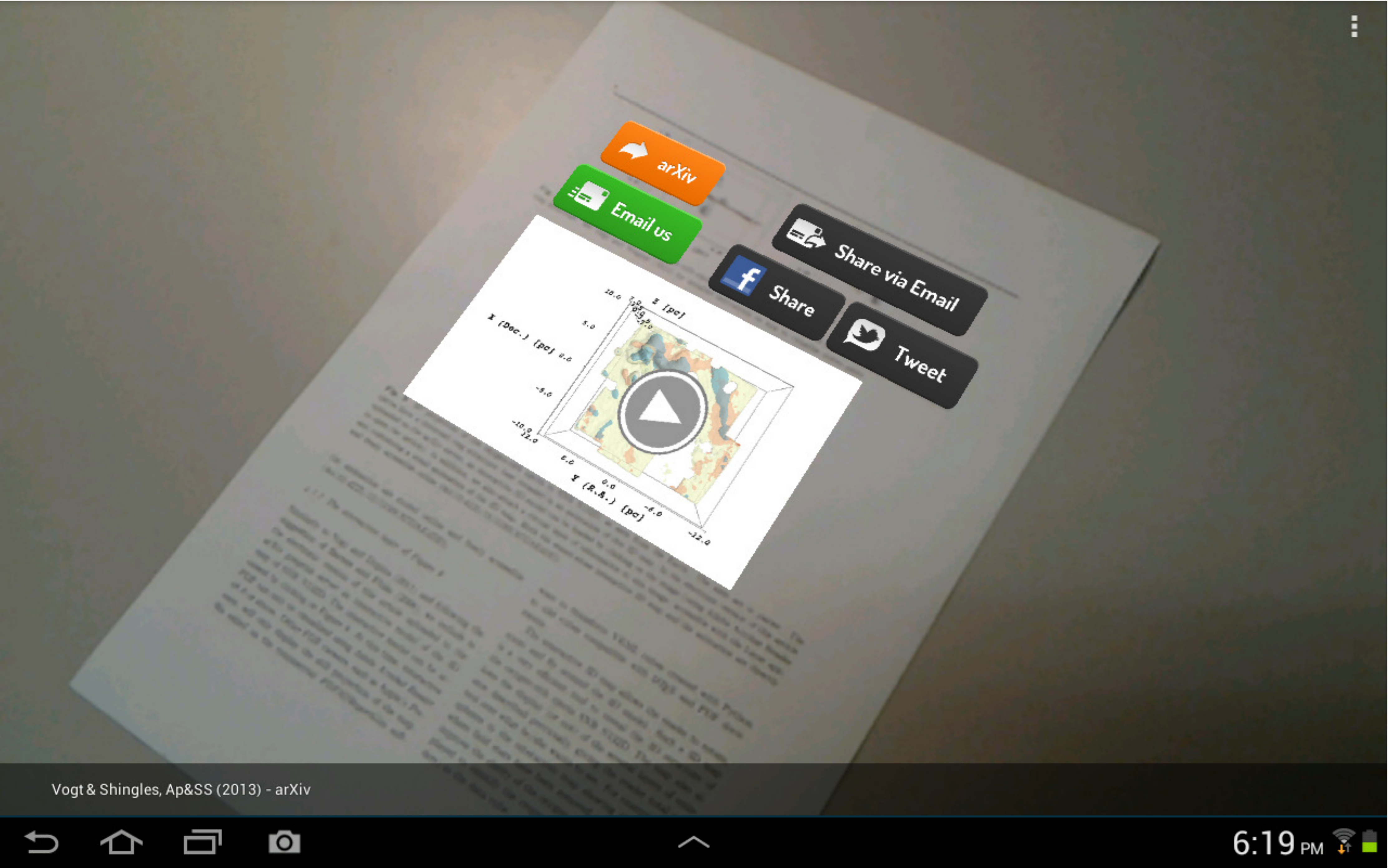}}
\caption{Screen capture of the top of page~\pageref{fig:n132d} seen through the Layar app. Clicking the floating image overlaid over Figure~\ref{fig:n132d} will start the animation. Additional buttons allowing the user to send us an email or share the article on social media are also included as additional examples of the wide variety of content that can be stored in a virtual layer of information.}\label{fig:layar}
\end{figure}

We attach to page~\pageref{fig:n132d} of this article a virtual layer containing an animation of the 3D map of SNR N132D. This virtual layer is accessible from both the printed and the electronic version of this article via the Layar app installed on a smartphone or tablet. In Figure~\ref{fig:layar}, we show a screen capture of the top of page~\pageref{fig:n132d} as seen through the Layar app. 

The animation starts from a view of the 3D map as seen from the Earth, rotates 90 degrees to place the reader in the plane of the sky, rotates a further 90 degrees to the left, and returns to the original point of view in the same (mirrored) way. Similar to the interactive 3D map, this animation allows the reader to clearly see where the oxygen-rich ejecta are located. 

Accessing the AR video with the Layar app is fast. On our iPhone 4S connected to the ANU wireless network, the AR layer is found within $\sim$4 seconds, and the video can start playing within an additional 2 seconds. The AR layer is not restricted to videos. For example, the AR layer attached to page~\pageref{fig:n132d} also contains 5 clickable buttons, allowing a) to contact us via email, b) to access the NASA/ADS entry for this article, and c) to share this article on social media or via email. Not included in the present example, additional material types, such as 3D models (similarly to the NASA Spacecraft 3D app) or sounds can also be added to the virtual layer of information.

The AR layer attached to this article was created with \emph{Layar Creator}. This online tool allows to easily upload pictures to the Layar database and attach virtual content in various formats. In its present form, Layar Creator requires registration but can be used for free. At the time of publication of this article, publishing augmented documents (i.e. adding a page to the Layar server) can be done for free, but Layar will include an advertisement banner at the bottom of the virtual layer of information. For this article, the virtual layer of information attached to page~\pageref{fig:n132d} was published advertisment-free, for a cost of 15 euros (rate of May 2013). There exists several alternatives to Layar Creator to publish content on the Layar server. These tools usually offer more freedom, but are also more complicated to set up. Extensive, detailed information on Layar is available on the app website, which we refer the reader to for more details\footnote{\href{http://layar.com/documentation/browser/layar-platform-overview/}{http://layar.com/documentation/browser/layar-platform-overview/}}.

We deliberately do not describe the Layar app in detail in this article, beyond the general principle that it uses image recognition technology to identify and overlay virtual datasets over physical content. Layar is easy to use and perfect for a first demonstration of AR at the research level in astrophysics in the form of Augmented Articles. Yet, for reasons discussed in Section~\ref{sec:stability}, the Layar app in its present form may not be ideal to be used extensively and on a regular basis by astronomers.

\section{Future trends for AR in astrophysics}\label{sec:future}

In the previous two sections, we have described two distinct examples of potential AR applications for astrophysical research. These two examples, guided by the characteristics and capabilities of different commercial softwares currently available, are designed to provide an overview of the potential of AR, and allow astronomers to directly experiment with this technology. The examples of Augmented Posters and Augmented Articles have been developed as reasonable applications of AR for the field of astrophysics. They should not be seen as the only possible applications of AR in this field. 

Predicting the future of AR is clearly premature at this stage, and strongly dependant on technological developments and the evolving needs of the end users. As illustrated in Sections~\ref{Sec:posters} and \ref{sec:papers}, the virtual layer of information associated with a a given physical target can host a wide variety of data formats: URLs, images, sounds, movies and 3D structures. Among these, 3D structures are clearly the most challenging to implement at the moment, due to a lack of software support (such 3D structures are also currently hard to produce in the first instance, but this discussion is outside the scope of this article). It appears rather unlikely that 3D structures will benefit most from AR technology in the short term, especially since the technology of interactive PDF is becoming more common and easy to use and implement in scientific journals. 

None of the other compatible data formats (images, videos, sounds) suffer from similar complex creation processes as 3D structures. As we have demonstrated, they can already be relatively easily implemented in AR applications that are currently available. Of these data formats, movies (or animation sequences) are possibly the most promising type of content for scientific AR applications. Movies are globally well supported by smartphones and other tablets. If they can be easily shared within the scientific community, they have the potential to strongly impact the quality and efficiency of the communication of scientific results, for example with theoretical simulations. Another possible - and as of now unique - expansion direction for AR is linked to the evolution of the role that social media plays in scientific research. The capability to directly and rapidly link and share scientific data and publications on social media, and that AR can offer (as illustrated in Section~\ref{sec:papers}), may be a powerful expansion driver for AR in astrophysical research, at least in the short term.

\section{Limitations of AR in astrophysical research}\label{sec:limitations}
General limitations of AR, such as social response, depth perception, or tracking techniques have already been discussed in the literature \citep[e.g.][]{Zhou08,Krevelen10}. In this section, we list and discuss several additional limitations, more specific to the field of astrophysics and science in general, that might impact the development and wide-spread usage of AR by researchers: complexity, accessibility and long-term stability.

\subsection{Complexity}
Until recently, implementing an AR system was a complex task requiring technical knowledge and dedicated efforts. In this article, we have demonstrated that this is not the case anymore. The identification of the commercial potential of AR, especially via smartphones and tablets, is giving rise to an increasing number of AR tools and apps. New apps, such as Shortcut or Layar, do not require specific knowledge to be implemented. AR is developing into a mature technology that can be easily used and implemented by non-experts \citep[e.g.][]{Vaughan09}. The range of available applications, relatively restricted at the time of publication of this article, can be expected to grow steadily. We strongly believe that astronomers, and scientists in general, can benefit from these developments. Complexity of AR is being taken away from the end user product, and is gradually less likely to slow down or stop the expansion of AR in the field of astrophysics, and science in general. We therefore believe that at this stage, the complexity associated with AR applications is not (anymore) the biggest limitation affecting their possible expansion to the field of astrophysics.

\subsection{Accessibility}

Any AR application will solely target smartphone and tablet users. This fact can be misinterpreted as discrimination from AR applications which do not address all scientists equally. We stress here that AR will never (or at least in the foreseeable future) replace physical content. In the examples we presented in this article, posters must be designed as a complete and understandable structure (and similarly for scientific articles). AR provides an alternative solution to easily share additional material, such as videos for example. Non-smartphone users can always access such videos via the internet with a computer - AR only simplifies the search process for scientists with a smartphone and allows them to access and visualize the additional material on-the-spot, potentially faster and more easily. With an ever increasing proportion of the population (and scientists) owning a smartphone or a tablet, AR has the potential to profoundly improve how complex data sets (and scientific results) are being shared within the community. In our opinion, AR applications should be seen (at this stage) as a possible complement to current standard data sharing methods, and not as a replacement. 

\subsection{Long-term stability}\label{sec:stability}
The issue of long-term stability and availability of AR material is a serious concern. The two AR examples implemented in this article rely on commercial products. There is no guarantee that the companies providing the Shortcut and Layar apps will still exist in 2, 10, or 50 years from the publication date of this article. There exists no guarantee either that future development of the Shortcut and Layar apps will ensure backwards compatibility with this article. The uncertainty of the long-term stability of AR applications is a strong limitation to their wide implementation in scientific publications. As we mentioned previously, AR only facilitates access to additional datasets that are not compatible with printed material. Measures should therefore be taken to ensure that this material remains accessible to the readers of the article in the far future. The long-term stability of the system and accessibility of the additional data is a (currently) significant shortcoming of AR compared to interactive PDF, for which the additional dataset can be stored directly with the support document.

We believe that a widespread usage of AR in astrophysical research would require the creation of a dedicated AR platform. This platform, created for scientists, would have long-term stability as a core principle. Not only would it ensure a long-term stability of the associated software, but it should also provide long-term storage solutions for any material linked to scientific documents. Such a dedicated tool would provide a unique, global service available to all scientists and journals. This platform would also ensure that published AR products are advertisement free. The associated app should be free to download, and available on all common smartphones and tablets. The running costs associated with data storage could be (for example) covered by charging a nominal (one-time) publication fee for virtual datasets, as well as (potentially) by receiving some financial support from participating scientific journals. 

Because the long-term stability requirement associated with scientific implementations of AR is not a feature shared by typical commercial AR applications, it appears unlikely (although not impossible) that a non-scientific organisation would agree to set up one such system. Most likely, the creation of a global AR platform for scientists will require a global effort and collaboration between researchers and publishers.

\section{Conclusion}\label{Sec:summary}

Using two distinct experiments, we have illustrated some of the potential of AR for astrophysical research, and science in general. AR is an emerging concept, but it is now transitioning to a more firmly established technology. We put forth the view that this technology may have a great but currently under-exploited potential in astrophysics. The key feature of this article is to enable its readers to directly experiment with AR. By doing so, this article allows astronomers to discover some of the potential of AR for the field of astrophysics, think about possible applications of this technology at the research level, and forge their own opinion about AR - not based on pre-conceived ideas, but rather on direct experimentation.

By implementing Augmented Poster sessions at the ASA AGM 2012 in Sydney, we have found evidence of some interest towards AR from Australian-based astronomers who experimented with this technology. Subsequent implementations of Augmented Poster sessions are required to clarify the exact amount of interest, and whether this precise AR application is indeed worth pursuing or not. 

In addition to our conference poster, attached in Appendix~\ref{app:poster} and compatible with the Shortcut app, we used the 3D map of the oxygen-rich ejecta in SNR N132D to illustrate the concept of Augmented Articles. Complementary to an interactive 3D map of SNR N132D accessible (with a computer) in the electronic version of this article, we attach a virtual layer of information to page~\ref{fig:n132d} in this article. This virtual layer (containing an animation of the 3D map of SNR N132D) can be accessed using the dedicated Layar app on a smartphone or tablet. This Augmented Article, the first of its kind, illustrates one possible application of AR in astrophysics at the research level. The exact form that AR may take for the field of astrophysics is uncertain. Movies, animations, and interactions with social media are our favoured directions for the expansion of AR in astrophysical research. In these domains, AR may be very complementary to alternative technologies, such as interactive PDF.

The main limitation of AR in astrophysics is related to the long-term stability and backwards compatibility of both the apps required to access virtual layers of information and the content of these information layers themselves. We advocate for the creation of a new AR service dedicated to publishing scientific AR material. An AR platform for scientists would ensure long-term stability and compatibility (for example of Augmented Articles), and may very strongly influence the future of AR at the research level in astrophysics, and science in general. It is unlikely that a non-scientific organisation would provide such a service. Cooperation between scientists and journals appears to be required to create such a system. In turn, such a global effort will take place only when scientists will have thought about the various possible applications of AR, clarified the exact potential of this technology, and decided whether it is a technology worth pursuing. Additional exploration and test-implementations of AR in the field of astrophysics at the research level are required, if astronomers are to decide on the usefulness of AR based on experimentation, rather than on pre-conceived ideas.

\acknowledgments

We thank Bill Roberts and the IT team at the Research School of Astronomy and Astrophysics (ANU) for their help in setting up the PDF3DReportGen software on the school servers. We also thank the anonymous referee for a very constructive review. We are grateful to the Kooaba team for allowing us to use their technology at the ASA AGM 2012. We thank the ASA AGM 2012 Local Organising Committee for their support in implementing Augmented Poster sessions during the conference. This research has made use of NASA's Astrophysics Data System. 

\appendix
\section{Vogt \& Shingles Interactive Poster presented at the ASA AGM 2012}\label{app:poster}
\centerline{\includegraphics[scale=0.75]{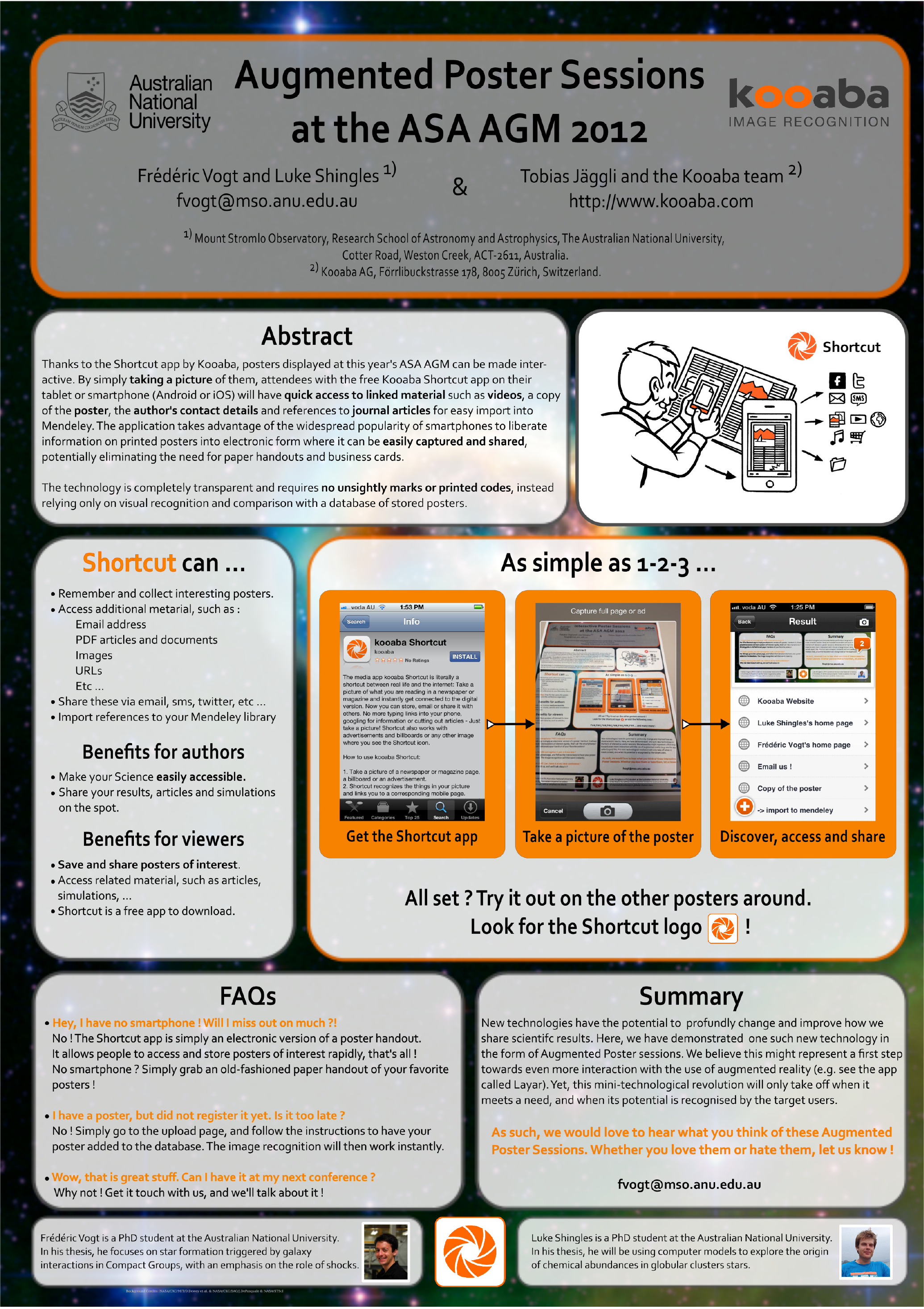}}

\newpage

\section{ASA AGM 2012 Feedback Survey on Augmented Posters}\label{App:survey}

The original survey was hosted online and could be accessed freely, without any registration required. The access link was circulated in an email sent to all the conference attendees shortly after the end of the conference. The 9 questions (and available answers to choose from) are listed below.

\begin{enumerate}
\item Did you attend this year's ASA AGM in Sydney ?
	\begin{itemize}
	\item Yes
	\item No
	\end{itemize}
\item Did you bring a poster with you ?
	\begin{itemize}
	\item Yes
	\item No
	\end{itemize}
\item Did you upload your poster on the online database to make it interactive ?
	\begin{itemize}
	\item Yes
	\item No
	\item I just told you I had no poster ...
	\end{itemize}
\item Did you know that the poster sessions were interactive (i.e that you could access related information by taking pictures of posters of interest) ?
	\begin{itemize}
	\item Yes, I knew it before the conference started.
	\item Yes, but I only found out after the conference started.
	\item No 
	\end{itemize}
\item Did you try to use the Shortcut app during the poster sessions ?
	\begin{itemize}
	\item Yes
	\item No, because I did not want to/ was not interested.
	\item No, because I don't have a smartphone.
	\item No, because ... [\textit{this option required to specify a reason}]
	\end{itemize}
\item What did you think of these interactive poster sessions ?
	\begin{itemize}
	\item Useless	
	\item Not terribly useful	
	\item They have some potential	
	\item Somewhat useful	
	\item Extremely useful
	\end{itemize}
\item Would you be interested to see this technology implemented in other conferences, such as the ASA AGM 2013 for example ?
	\begin{itemize}
	\item Yes
	\item No
	\item I don't care
	\end{itemize}
\item Do you have any specific comment ? [\textit{Answer to this question was free and optional}]
\item Final question ! You are ...
	\begin{itemize}
	\item an undergraduate student
	\item a PhD student
	\item a postdoc
	\item a professor
	\item something else
	\end{itemize}
\end{enumerate}

\end{document}